\documentclass[%
preprint,
 amsmath,amssymb,
 aps,
]{revtex4}

\usepackage{graphicx}
\usepackage{bm}

\begin{document}

\title{Mapping the XY Hamiltonian onto a Network of Coupled Lasers}

\author{Mostafa Honari-Latifpour}
\author{Mohammad-Ali Miri}%
 \email{mmiri@qc.cuny.edu}
\affiliation{%
 Department of Physics, Queens College of the City University of New York, Queens, New York 11367 USA\\
 Physics Program, The Graduate Center, City University of New York, New York, New York 10016, USA
}%

\date{August 25, 2020}%

\begin{abstract}

In recent years there has been a growing interest in the physical implementation of classical spin models through networks of optical oscillators. However, a key missing step in this mapping is to formally prove that the dynamics of such a nonlinear dynamical system is toward minimizing a global cost function which is equivalent with the spin model Hamiltonian. Here, we introduce a minimal dynamical model for a network of dissipatively coupled optical oscillators and prove that the dynamics of such a system is governed by a Lyapunov function that serves as a cost function for the system. This cost function is in general a function of both phases and intensities of the oscillators and depends strongly on the pump parameter. In case of bipartite network topologies, the amplitudes of the oscillators become identical in the steady state and the cost function reduces to the XY Hamiltonian. In the general case for non-trivial network topologies, however, the cost function approaches the XY Hamiltonian only in the strong pump limit. We show that by adiabatically tuning the pump parameter, the network can largely avoid trapping into the local minima of the governing cost function and stabilize into the ground state of the associated XY Hamiltonian. These results show the great potential of laser networks for unconventional computing.

\end{abstract}

\maketitle


\noindent
Classical spin models are widely utilized in statistical mechanics and condensed matter physics for exploring critical phenomena and phase transitions in magnetic materials \cite{baxter2016exactly,edwards1975theory}. Beyond their original realm, these models have been also applied to investigate a wide range of complex phenomena, such as collective behavior of neural networks \cite{hopfield1982neural} and protein folding \cite{bryngelson1987spin}. In addition, they have inspired efficient heuristics in combinatorial optimization, which makes them an attractive alternative to conventional methods for solving computationally hard problems \cite{kirkpatrick1983optimization,fu1986application}. Consequently, the possibility of realizing an analog spin lattice model is of great interest.

Recently, there has been a growing interest in emulating spin models with nonlinear driven-damped optical systems \cite{Utsunomiya:11, wang2013coherent,marandi2014network,nixon2013observing, mcmahon2016fully, inagaki2016coherent, takeda2017boltzmann, berloff2017realizing, PhysRevLett.121.235302, parto2020realizing, pierangeli2019large, bohm2019poor, Honari_2020}. In particular, networks of coherently coupled degenerate optical parametric oscillators were used for implementing a binary spin system in analogy with the Ising model and utilized for solving NP-hard problems \cite{wang2013coherent, marandi2014network}. In addition, the phase pattern of large arrays of dissipatively coupled solid-state lasers were shown to be analogous to arrangement of spins governed by the XY Hamiltonian \cite{nixon2013observing}. Similar behavior was also observed in the polarization states of nano-laser arrays \cite{parto2020realizing}. Furthermore, networks of parametric three-photon down-conversion oscillators have been suggested for implementing a three-state Potts machine \cite{Honari_2020}.

In these contexts, a network of interacting optical oscillators are brought into a phase-locked state, where the intensities tend to be uniform across the network, while the phases reveal striking patterns \cite{marandi2014network, nixon2013observing}. In case of coupled lasers, assuming that the intensities of all lasers are equal, the phases are shown to be governed by an energy landscape function which turns out to be identical with an anti-ferromagnetic XY Hamiltonian \cite{nixon2013observing}. However, it remains to analytically investigate the assumption of uniform equilibrium intensity which is critical to a faithful mapping of the XY Hamiltonian onto a network of lasers. Consequently, it is of great interest to derive an exact cost function for the laser network which in general involves both the intensity and phase degrees of freedom. In addition, to the best of our knowledge, there is no formal proof of the evolution of the above-mentioned machines toward a state with globally minimum modal loss, as suggested in previous works \cite{wang2013coherent, marandi2014network}. Finally, it is critical to investigate the stability of such highly nonlinear systems in order to ensure their proper operation in presence of inevitable imperfections.

\begin{figure}[htbp]
\includegraphics[width=8.6cm]{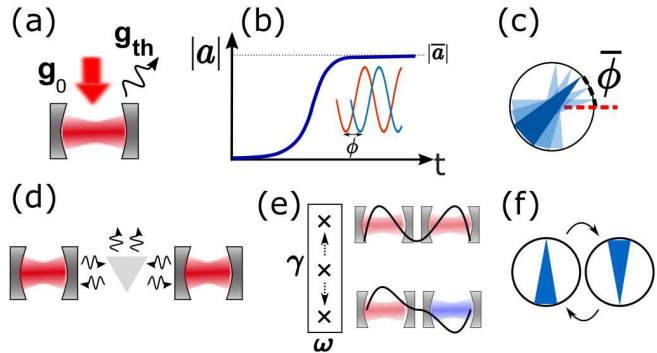}
\caption{A schematic illustration of the mapping of the XY model onto lasers. (a) A single laser. (b) Temporal evolution of the complex modal field amplitude of a laser. (c) The steady state of the complex field amplitude is shown by a vector in the 2D plane, while its magnitude and angle respectively represent the oscillation intensity and phase. (d) An arrangement of two lasers interacting through dissipation into a scattering channel. (e) Splitting of the linear eigenfrequencies of the system along the imaginary axis as a result of their dissipative coupling. (f) The oscillation of the coupled lasers into a preferred phase-locked state with $\pi$ phase contrast resembles the ground state of anti-ferromagnetic system.}
\label{fig1}
\end{figure}

In this Letter, by introducing an integrable model, we systematically explore the problem of mapping the classical XY model onto networks of optical oscillators with amplitude and phase degrees of freedom. As a building block of our model, we consider a single-mode laser as depicted schematically in Fig.~\ref{fig1}(a). In a semi-classical treatment and by adiabatic elimination of the atomic variables, the laser field is described with a nonlinear oscillator model \cite{lamb1964theory, mandel1995optical}. The evolution equation of such an oscillator is $\dot{a}(t)=(-i\omega_0 + g_0 - g_{\textup{th}} - g_s |a|^2) a$, where, $a$ represents the complex field amplitude, $\omega_0$ is the oscillation frequency, $g_{\textup{th}}$ is the total laser losses, $g_0$ is the linear gain and  $g_s$ is the gain saturation coefficient. 
This equation admits the solution $a=\sqrt{I(t)}\exp(-i\omega_0 t + \bar{\phi})$, where, $\bar{\phi}$ is an arbitrary phase, and $I^{-1}(t)={\bar{I}}^{-1}+(I_0^{-1}-{\bar{I}}^{-1})\exp{[-2(g_0-g_{\textup{th}})t]}$, where, $\bar{I}=(g_0-g_{\textup{th}})/g_s$ is the steady-state intensity, and $I_0$ is the initial intensity. According to this relation for $g_0-g_{\textup{th}}>0$, the field builds up to a steady-state amplitude $|\bar{a}|=\sqrt{(g_0-g_{\textup{th}})/g_s}$, while it exhibits an arbitrary phase $0<\bar{\phi}<2\pi$ (Fig.~\ref{fig1}(b)). As depicted in Fig.~\ref{fig1}(c), the steady-state complex field can be described with a vector in the 2D plane such that its magnitude and angle represent $|\bar{a}|$ and $\bar{\phi}$, respectively.

\begin{figure}[htbp]
\includegraphics[width=8.6cm]{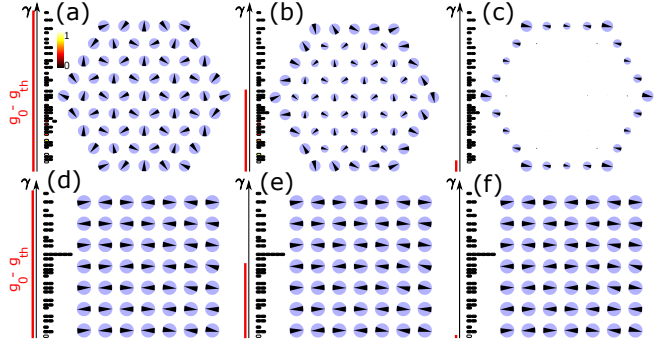}
\caption{The equilibrium state of a network of dissipatively coupled oscillators with triangular (top) and rectangular (bottom) lattice topologies and for different levels of the linear gain parameter. In each panel, the eigenvalues of the $Q$ matrix, $\gamma_1,\cdots,\gamma_N$, are sorted and shown with markers as a ladder along the vertical axis, while the red bars show the level of the linear differential gain $g_0-g_\textup{th}$. The projection of the equilibrium state on the eigenvectors of the $Q$ matrix is color-coded on the associated eigenvalue markers. In these simulations, the coupling is assumed to be limited to nearest neighbors with uniform strength. The differential gain and the gain saturation coefficients are chosen such that $\bar{I}=0.98$ (a), $0.96$ (b), and $0.91$ (c), $0.98$(d), $0.97$ (e), and $0.57$ (f).}
\label{fig2}
\end{figure}

Although the steady-state phase of a single oscillator may not be of particular interest, it finds meaning when two such oscillators are coherently coupled. In this case, the two oscillators come to a phase-locking even in presence of tolerable initial frequency detunings \cite{spencer1972theory}. The synchronization process becomes particularly appealing when the two oscillators are coupled dissipatively as illustrated in Fig.~\ref{fig1}(d) \cite{ding2019dispersive}. The interesting property of such a dissipative interaction is the coherent superposition of the radiative fields from the two resonators which creates a contrast in the level of radiation losses for the two eigenmodes of the system (Fig.~\ref{fig1}(e)). Therefore, when the gain is turned on, the system tends to evolve toward the eigenmode with minimum leakage. Quite interestingly, in the steady state, the two oscillators reach the same intensity, while the phase contrast is close to $\pi$ \cite{ding2019dispersive} (Fig.~\ref{fig2}(e)).

This process can be viewed as a search toward an optimal state in the phase space of the system. Assuming that the two oscillators are arranged such that they equally radiate in the leakage channel, the rate of energy dissipation is $P_\textrm{diss} \propto \kappa_{12} \left| |a_1| e^{i\phi_1} + |a_2| e^{i\phi_2} \right|^2$, where $\kappa_{12}$ represents the dissipative coupling rate. This latter relation is of course minimized for the trivial choice of zero oscillator amplitudes. However, one should consider the constraint imposed on the amplitude of each oscillator through the pump. Assuming that the two oscillators reach the same steady-state intensity $|a_{1,2}|=|\bar{a}|$, the dissipated power simplifies to $P_\textrm{diss} \propto \kappa_{12} |\bar{a}| \left[1 + \cos(\phi_1-\phi_2) \right]$, which, is identical to the classical XY Hamiltonian for a lattice with two spins. However, this mathematical analogy is built on assuming equal steady-state intensities for the two oscillators.

Considering a network of $N$ dissipatively coupled identical oscillators, by using a gauge transformation $a_m \rightarrow a_m e^{-i\omega_0 t}$, the time evolution equation governing the complex modal amplitude of the $m$'th oscillator can be written as:
\begin{equation}
\label{eq1}
    \dot{a}_m=\left(g_0-g_{\textup{th}} - g_s|a_m|^2\right)a_m - \textstyle\sum\limits_{n\neq m} \kappa_{mn} (a_m + a_n).
\end{equation}
Here, $\kappa_{mn}$ is the coupling coefficients between the $m$'th and $n$'th lasers. The diagonal element appearing in the summation represents the external losses due to dissipative coupling, thus, the total loss of the $m$'th resonator is the sum of its intrinsic and external losses: $g_{\textup{th}}+\sum_{n\neq m} \kappa_{mn}$. In writing equations (\ref{eq1}), we assume that the dissipative coupling occurs only pairwise and through decaying into a common dissipation channel. Furthermore, the coupling coefficients are assumed to be non-negative $\kappa_{mn} \geq 0$, which is equivalent to considering only in-phase addition of the decaying fields from the two resonators. In addition, the coupling coefficients are assumed to be symmetric, i.e., $\kappa_{mn}=\kappa_{nm}$.


The system of equations~(\ref{eq1}) can be described through a cost function $F(a_1,a_1^*,\cdots,a_N,a_N^*)$ such that $\dot{a}_m=-\partial F / \partial a_m^*$ and $\dot{a}_m^*=-\partial F / \partial a_m$. By direct integration of Eq.~(\ref{eq1}) and by addition of a suitable constant, $F$ is found to be
\begin{equation}
\label{eq3}
    F=\tfrac{g_s}{2}\textstyle\sum_{m}^{}(|a_m|^2 - |\bar{a}|^2)^2 + \frac{1}{2}\textstyle\sum_{m,n}^{} \kappa_{mn} \left| a_m + a_n \right|^2
\end{equation}
where, $|\bar{a}|^2=(g_0-g_{\textup{th}})/g_s$, is the steady-state intensity of a single oscillator.

It is obvious that $F$ is locally positive-semidefinite, while its total time derivative along the trajectories of Eq.~(\ref{eq1}) is $dF/dt = -2 \sum_{m}^{}|\dot{a}_m|^2 $, which is locally negative-semidefinite. These conditions ensure the evolution of the system from a given point in the phase space toward a state of equilibrium that minimizes $F$ (locally or globally) \cite{hahn1963theory}. The existence of the functional $F$ with the properties mentioned above, along with the fact that it is radially unbounded, guarantees local stability of the equilibrium states of the system.

By rewriting Eq.~(\ref{eq3}) using the intensity and phase representation as $F=\frac{g_s}{2}\sum_{m}^{}(I_m - \bar{I})^2 + \frac{1}{2}\sum_{m,n}^{} \kappa_{mn} \left[I_m+I_n + 2\sqrt{I_m I_n} \cos{(\phi_m-\phi_n)} \right]$, it becomes clear that the XY Hamiltonian is embedded in this cost function. In order to interpret the cost function and to investigate its relation with modal losses, first we cast the dynamical equations (\ref{eq1}) in a matrix form as follows:
\begin{equation}
\label{eq4}
\dot{\mathbf{a}}=\mathbf{f}(\mathbf{a})-Q\mathbf{a},
\end{equation}
Here, $\mathbf{a}=[a_1,\cdots,a_N]^t$, and $\mathbf{f}(\mathbf{a})=[f_1(a_1),\cdots,f_N(a_N)]^t$, where $f_m(a_m)=(g_0-g_{\textup{th}}-g_s|a_m|^2)a_m$, and $Q$ is a signless Laplacian matrix with off-diagonal elements $q_{mn}=\kappa_{mn}$ and diagonal elements $q_{mm}=\sum_{n\neq m}{}\kappa_{mn}$.

In this representation, the dynamical equations can be decomposed into a nonlinear diagonal term $\mathbf{f}(\mathbf{a})$ and an interaction term $-Q\mathbf{a}$. 
Apart from the intrinsic loss $g_{\textup{th}}$, the eigenmodes of the interaction term, $Q\mathbf{v}_i=\gamma_i \mathbf{v}_i ~ ; ~ i=1, \cdots ,N$, represent the linear modal losses of the network. Given that $Q$ is a real symmetric matrix, its eigenvalues are real and can be sorted as $\gamma_1 \leq \gamma_2 \leq \cdots \leq \gamma_N$. One can define a loss functional in form of a Rayleigh quotient:
\begin{equation}
\label{eq5}
    \Gamma[\mathbf{a},\mathbf{a}^*] = \frac{\mathbf{a}^{\dagger} Q \mathbf{a}}{\mathbf{a}^{\dagger} \mathbf{a}}
\end{equation}
which, its minimum value is the smallest eigenvalue of the matrix $Q$ and that occurs at the corresponding eigenvector. The cost function of Eq.~(\ref{eq3}) is cast in the matrix form as follows:
\begin{equation}
\label{eq6}
    F[\mathbf{a},\mathbf{a}^*] = \frac{g_s}{2} (\mathbf{I} - \mathbf{\bar{I}})^{\dagger}(\mathbf{I} - \mathbf{\bar{I}}) + \frac{1}{2}\mathbf{a}^{\dagger} Q \mathbf{a}
\end{equation}
where, $\mathbf{I}=[|a_1|^2, \cdots , |a_N|^2]^t$ and $\mathbf{\bar{I}}=|\bar{a}|^2[1,\cdots,1]^t$.

It is straightforward to show that both functionals of Eqs.~(\ref{eq5},\ref{eq6}) become zero when the underlying network graph is bipartite, i.e., its nodes can be separated into two disjoint sets such that all links are located between the two sets. In fact, the smallest eigenvalue of the signless Laplacian matrix of a graph is zero if and only if it is bipartite \cite{cvetkovic2007signless}. In addition, the associated eigenvector takes values of $+1$ and $-1$ on nodes located in the two disjoint parts of the network. Therefore, for a bipartite network, the eigenvector associated with the smallest eigenvalue of the $Q$-matrix is an equilibrium state of the oscillator network with minimum cost ($F=0$).

The conditions for reaching an equilibrium state with uniform intensity can be explored by directly enforcing the ansatz of $|a_m(t)|=|a_{\textup{ss}}|$ for $m=1,\cdots,N$, in the dynamical equations~(\ref{eq4}), which results in the algebraic equation $Q \mathbf{a}=(g_0-g_{\textup{th}}-g_s|a_{\textup{ss}}|^2) \mathbf{a}$, under the constraint of $|a_1|=\cdots=|a_N|$. In case of the bipartite graphs, the answer becomes trivial since the network stabilizes to the eigenvector associated with the smallest eigenvalue, thus $(g_0-g_{\textup{th}}-g_s|a_{\textup{ss}}|^2)=0$. However, there is no simple answer to the question of the existence of an eigenvector with uniform intensity for the $Q$ matrix associated with a general network, except for special cases such as an all-to-all connected graph. Nonetheless, the cost function provides insight into the equilibrium intensity pattern of general networks as we discuss in the following.

According to Eqs.~(\ref{eq3},\ref{eq6}), the cost function is the sum of a self-oscillation term and an interaction term, where both contributions are non-negative. Considering these two terms individually, the first is minimized when all oscillators reach the same steady-state intensity $\bar{I}$ as in a single oscillator. The second term becomes zero for the trivial choice of $\mathbf{a}=\mathbf{0}$. On the other hand, minimizing the second term subject to finite intensities requires an optimal configuration of the phases. Therefore, the equilibrium state emerges as a result of a balance between two competing contributions in the cost function; the self-oscillation term that tends to adjust the intensities to a fixed value, and the interaction term that tends to reduce the intensities and simultaneously organize the phases.


The competition between the two terms of the cost function can be evaluated through the relative strength of the drive $g_0-g_\textup{th}$ versus the set of coupling coefficients $\{\kappa_{mn}\}$, which involves both the strength of the interactions and the network topology. To explore the role of the pump parameter, we compare two cases of a bipartite and a non-bipartite system with rectangular and triangular lattice topologies. Figure~\ref{fig2} depicts the steady-state pattern of the two lattices for three different levels of the pump parameter. In this figure, the eigenvalues $\gamma_1,\cdots,\gamma_N$, and the differential gain level $g_0-g_{\textup{th}}$, are respectively shown with ladder of markers and red bars along the vertical axes. In each case, the equilibrium state is projected on the associated eigenvectors of the coupling matrix $Q$ and the magnitude of the projection coefficients are color coded on eigenvalue markers. As clearly indicated in Figs.~\ref{fig2}(a-c), the non-bipartite lattice behaves completely different under different pump levels. In this case, for high gain levels the steady state approaches toward a uniform intensity pattern. By decreasing the gain, however, an intensity contrast appears between the bulk oscillators and those located on the edge. In case of the bipartite network, on the other hand, as shown in Figs.~\ref{fig2}(d-f), for all values of the pump parameter, the network stabilizes to the same pattern which is associated with the eigenstate with the lowest modal loss. To further explore these results, similar simulations were performed for all connected graph topologies with six nodes, involving 112 cases. The results are shown in the Supplementary Material, showing a consistent trend in all cases \cite{suppl}. These results indicate that in general the presence of odd cycles spoils the uniform equilibrium intensity pattern in the small gain limit.

The contrast in the steady-state intensity pattern of the system in the weak and strong pump regimes can be explained in terms of the cost function. In the small gain regime, $\bar{I}$ is small, thus the system affords to enforce zero intensity for some oscillators in favor of minimizing the second term of the cost function. In contrast, in the large gain regime, death of an oscillator will significantly increase the self-oscillation term. As a result, the steady state tends to approach a uniform intensity pattern while the phase pattern is organized such that the second term is minimized. Therefore, for the general case of a non-bipartite graph the mapping of the XY Hamiltonian onto the network of coupled oscillators becomes accurate in the strong pump regime. It is worth noting that for the triangular lattice discussed in Fig.~\ref{fig2}, in the weak pump limit the preferential death of oscillators happens for the bulk oscillators since they are coupled to more elements and thus their death lead to a greater reduction of the cost function. Here, the governing dynamical equations are in essence different from a recently demonstrated topological insulator laser, which is governed by the Haldane Hamiltonian \cite{harari2018topological, bandres2018topological}.
\begin{figure}
    \centering
    \includegraphics[width=8.6cm]{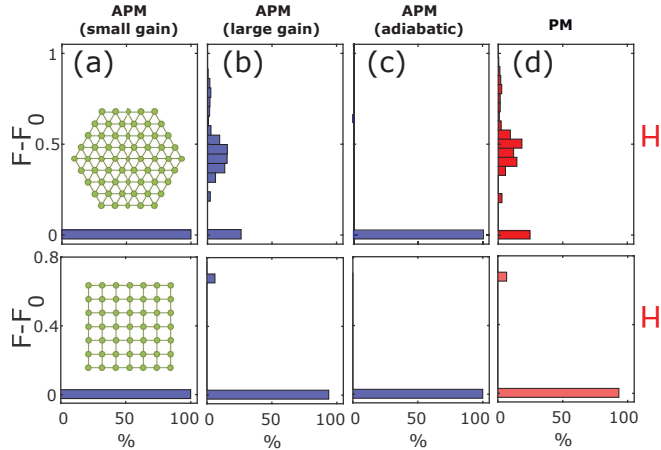}
    \caption{The distribution of the equilibrium state cost function $F$ associated with the amplitude-and-phase model (APM) of Eq.~(\ref{eq1}) (a-c) and the cost function $H$ of the phase model (PM) of Eq.~(\ref{eqK}) (d) for the triangular (top) and rectangular (bottom) lattices of Fig.~\ref{fig2}. For the triangular lattice, the gain is such that $\bar{I}=0.8960$ in (a), and $0.9979$ in (b), while for the rectangular lattice, this parameter is $0.0099$ and $0.9965$ in (a) and (b), respectively. In panel (c), the gain is linearly tuned such that $\bar{I}$ adiabatically increases from $0$ to $0.9979$ for the triangular lattice and from $0$ to $0.9965$ for the rectangular lattice. Each figure is obtained by $10,000$ simulations with a random ensemble of initial conditions; in (a-c), the initial amplitudes $|a_m(0)|$ are randomly selected from the range $ [0.01, 0.05]$, and in (a-d) the initial phases $\phi_m(0)$ are randomly selected from $[-\pi, \pi]$ with uniform probability.}
    \label{fig3}
\end{figure}

It is important to note that the steady-state patterns shown in Fig.~\ref{fig2} are global minima of the associated networks for the given gain levels. On the other hand, the cost function of Eq.~(\ref{eq3}) guarantees stability in a local sense. Thus, the attractor basin of an equilibrium point could be a finite region in the $2N$-dimensional phase space, and a perturbation can move the system from one equilibrium point to another. In order to investigate this aspect, we explored the equilibrium state statistics of the networks of Fig.~\ref{fig2} at different gain levels and for large ensembles of initial conditions. The results are shown in Figs.~\ref{fig3}(a,b) for two extreme cases of small and large gains, while additional cases are shown in the Supplementary Material \cite{suppl}. The results suggest that the non-bipartite network involves a more complex cost function with a larger number of local minima states. In addition, in both cases of bipartite and non-bipartite networks, the chances of trapping into the local minima increases for higher gain levels. The trapping of the network to local minima can be circumvented by gradually increasing the gain level as shown in Fig.~\ref{fig3}(c). In this manner, the cost function gradually deforms to the XY Hamiltonian, while its global minimum state adiabatically transforms into the ground state of the XY model.

According to the above discussion, by simulating the dynamical model of Eq.~(\ref{eq1}), one can find the ground state of the associated XY Hamiltonian $H=\sum_{m,n}\kappa_{mn} [1 + \cos(\phi_m - \phi_n)]$, which may generally involve many local minima. In order to show the performance of the dynamical model of Eq.~(\ref{eq1}) as an optimizer of the XY Hamiltonian, we compare it with a direct gradient-based optimization of the XY Hamiltonian, according to the dynamical model, $\dot{\phi}_m=-\partial H / \partial \phi_m$:
\begin{equation}
\label{eqK}
\dot{\phi}_m=-\sum_{n}\kappa_{mn} \sin(\phi_n - \phi_m).
\end{equation}
This is the well-known Kuramoto model on a graph with weights $-\kappa_{mn}$ \cite{nixon2013observing, acebron2005kuramoto}. Figure \ref{fig3}(d) depicts the distribution of the XY energy for the triangular and rectangular lattices of Fig.\ref{fig2} by simulating Eq.~(\ref{eqK}) for a large ensemble of initial conditions. The astonishing similarity of Figs.~\ref{fig3}(b,d) again indicates the equivalence of the cost function of the oscillator network in the large gain limit with the XY Hamiltonian. However, a comparison between Figs.~\ref{fig3}(c,d) reveals the superior performance of the dynamical model of Eq.~(\ref{eq1}) over that of Eq.~(\ref{eqK}) for globally minimizing the XY Hamiltonian. In this case, for $100\%$ of the simulation incidents both networks stabilized into their global minima, while for the phase model the success rate is around $30\%$ and $90\%$ for the triangular and rectangular networks, respectively. This is owing to the additional amplitude degree of freedom in Eq.~(\ref{eq1}), which allows for adiabatically deforming the associated cost function towards the XY Hamiltonian while avoiding the local minima. These results clearly indicate the potential of laser networks for unconventional computing applications.

In summary, by introducing an integrable model, we studied the dynamics of a network of dissipatively coupled lasers and its operation as a classical XY simulator. The governing cost function involves both amplitude and phase degrees of freedom and depends strongly on the gain parameter. For non-trivial network topologies, the mapping to the XY Hamiltonian becomes accurate only in the strong pump regime. In addition, we showed that adiabatic tuning of the pump parameter can greatly assist the network to avoid trapping into the local minima of the governing cost function to stabilize into the ground state of the associated XY Hamiltonian. These findings can serve as a key step in optical realization of spin lattices for unconventional computing.
\begin{acknowledgments}
The authors gratefully acknowledge useful discussions with Dr. Alireza Marandi.
\end{acknowledgments}


\begin{thebibliography}{10}

\bibitem{baxter2016exactly}
R.~J. Baxter, {\em Exactly solved models in statistical mechanics}.
\newblock Academic Press Limited, 1982.

\bibitem{edwards1975theory}
S.~F. Edwards and P.~W. Anderson, ``Theory of spin glasses,'' {\em J. Phy. F},
  vol.~5, pp.~965--974, may 1975.

\bibitem{hopfield1982neural}
J.~J. Hopfield, ``Neural networks and physical systems with emergent collective
  computational abilities,'' {\em Proc. Natl. Acad. Sci. U.S.A.}, vol.~79,
  no.~8, pp.~2554--2558, 1982.

\bibitem{bryngelson1987spin}
J.~D. Bryngelson and P.~G. Wolynes, ``Spin glasses and the statistical
  mechanics of protein folding,'' {\em Proc. Natl. Acad. Sci. U.S.A.}, vol.~84,
  no.~21, pp.~7524--7528, 1987.

\bibitem{kirkpatrick1983optimization}
S.~Kirkpatrick, C.~D. Gelatt, and M.~P. Vecchi, ``Optimization by simulated
  annealing,'' {\em Science}, vol.~220, no.~4598, pp.~671--680, 1983.

\bibitem{fu1986application}
Y.~Fu and P.~W. Anderson, ``Application of statistical mechanics to
  {NP}-complete problems in combinatorial optimisation,'' {\em J. Phys. A},
  vol.~19, pp.~1605--1620, jun 1986.

\bibitem{Utsunomiya:11}
S.~Utsunomiya, K.~Takata, and Y.~Yamamoto, ``Mapping of ising models onto
  injection-locked laser systems,'' {\em Opt. Express}, vol.~19,
  pp.~18091--18108, Sep 2011.

\bibitem{wang2013coherent}
Z.~Wang, A.~Marandi, K.~Wen, R.~L. Byer, and Y.~Yamamoto, ``Coherent ising
  machine based on degenerate optical parametric oscillators,'' {\em Phys. Rev.
  A}, vol.~88, p.~063853, Dec 2013.

\bibitem{marandi2014network}
A.~Marandi, Z.~Wang, K.~Takata, R.~L. Byer, and Y.~Yamamoto, ``Network of
  time-multiplexed optical parametric oscillators as a coherent {Ising}
  machine,'' {\em Nature Photonics}, vol.~8, pp.~937--942, Dec. 2014.

\bibitem{nixon2013observing}
M.~Nixon, E.~Ronen, A.~A. Friesem, and N.~Davidson, ``Observing geometric
  frustration with thousands of coupled lasers,'' {\em Phys. Rev. Lett.},
  vol.~110, p.~184102, May 2013.

\bibitem{mcmahon2016fully}
P.~L. McMahon, A.~Marandi, Y.~Haribara, R.~Hamerly, C.~Langrock, S.~Tamate,
  T.~Inagaki, H.~Takesue, S.~Utsunomiya, K.~Aihara, R.~L. Byer, M.~M. Fejer,
  H.~Mabuchi, and Y.~Yamamoto, ``A fully programmable 100-spin coherent ising
  machine with all-to-all connections,'' {\em Science}, vol.~354, no.~6312,
  pp.~614--617, 2016.

\bibitem{inagaki2016coherent}
T.~Inagaki, Y.~Haribara, K.~Igarashi, T.~Sonobe, S.~Tamate, T.~Honjo,
  A.~Marandi, P.~L. McMahon, T.~Umeki, K.~Enbutsu, O.~Tadanaga, H.~Takenouchi,
  K.~Aihara, K.-i. Kawarabayashi, K.~Inoue, S.~Utsunomiya, and H.~Takesue, ``A
  coherent ising machine for 2000-node optimization problems,'' {\em Science},
  vol.~354, no.~6312, pp.~603--606, 2016.

\bibitem{takeda2017boltzmann}
Y.~Takeda, S.~Tamate, Y.~Yamamoto, H.~Takesue, T.~Inagaki, and S.~Utsunomiya,
  ``Boltzmann sampling for an xy model using a non-degenerate optical
  parametric oscillator network,'' {\em Quantum Science and Technology},
  vol.~3, no.~1, p.~014004, 2017.

\bibitem{berloff2017realizing}
N.~G. Berloff, M.~Silva, K.~Kalinin, A.~Askitopoulos, J.~D. Töpfer,
  P.~Cilibrizzi, W.~Langbein, and P.~G. Lagoudakis, ``Realizing the classical
  {XY} {Hamiltonian} in polariton simulators,'' {\em Nature Materials},
  vol.~16, pp.~1120--1126, Nov. 2017.

\bibitem{PhysRevLett.121.235302}
K.~P. Kalinin and N.~G. Berloff, ``Simulating ising and $n$-state planar potts
  models and external fields with nonequilibrium condensates,'' {\em Phys. Rev.
  Lett.}, vol.~121, p.~235302, Dec 2018.

\bibitem{parto2020realizing}
M.~Parto, W.~Hayenga, A.~Marandi, D.~N. Christodoulides, and M.~Khajavikhan,
  ``Realizing spin {Hamiltonians} in nanoscale active photonic lattices,'' {\em
  Nature Materials}, Mar. 2020.

\bibitem{pierangeli2019large}
D.~Pierangeli, G.~Marcucci, and C.~Conti, ``Large-scale photonic ising machine
  by spatial light modulation,'' {\em Phys. Rev. Lett.}, vol.~122, p.~213902,
  May 2019.

\bibitem{bohm2019poor}
F.~B{\"o}hm, G.~Verschaffelt, and G.~Van~der Sande, ``A poor man’s coherent
  ising machine based on opto-electronic feedback systems for solving
  optimization problems,'' {\em Nat. Commun.}, vol.~10, no.~1, pp.~1--9, 2019.

\bibitem{Honari_2020}
M.~Honari-Latifpour and M.-A. Miri, ``Optical potts machine through networks of
  three-photon down-conversion oscillators,'' {\em Nanophotonics}, no.~0,
  p.~20200256, 2020.

\bibitem{lamb1964theory}
W.~E. Lamb, ``Theory of an optical maser,'' {\em Phys. Rev.}, vol.~134,
  pp.~A1429--A1450, Jun 1964.

\bibitem{mandel1995optical}
L.~Mandel and E.~Wolf, {\em Optical coherence and quantum optics}.
\newblock Cambridge University Press, 1995.

\bibitem{spencer1972theory}
M.~B. Spencer and W.~E. Lamb~Jr, ``Theory of two coupled lasers,'' {\em
  Physical Review A}, vol.~5, no.~2, p.~893, 1972.

\bibitem{ding2019dispersive}
J.~Ding, I.~Belykh, A.~Marandi, and M.-A. Miri, ``Dispersive versus dissipative
  coupling for frequency synchronization in lasers,'' {\em Phys. Rev. Applied},
  vol.~12, p.~054039, Nov 2019.

\bibitem{hahn1963theory}
W.~Hahn, H.~H. Hosenthien, and H.~Lehnigk, {\em Theory and application of
  Liapunov's direct method}.
\newblock Prentice-Hall Englewood Cliffs, NJ, 1963.

\bibitem{cvetkovic2007signless}
D.~Cvetkovi{\'c}, P.~Rowlinson, and S.~K. Simi{\'c}, ``Signless laplacians of
  finite graphs,'' {\em Linear Algebra Its Appl.}, vol.~423, no.~1,
  pp.~155--171, 2007.

\bibitem{suppl}
``See supplemental material.''

\bibitem{harari2018topological}
G.~Harari, M.~A. Bandres, Y.~Lumer, M.~C. Rechtsman, Y.~D. Chong,
  M.~Khajavikhan, D.~N. Christodoulides, and M.~Segev, ``Topological insulator
  laser: theory,'' {\em Science}, vol.~359, no.~6381, p.~eaar4003, 2018.

\bibitem{bandres2018topological}
M.~A. Bandres, S.~Wittek, G.~Harari, M.~Parto, J.~Ren, M.~Segev, D.~N.
  Christodoulides, and M.~Khajavikhan, ``Topological insulator laser:
  Experiments,'' {\em Science}, vol.~359, no.~6381, p.~eaar4005, 2018.

\bibitem{acebron2005kuramoto}
J.~A. Acebr\'on, L.~L. Bonilla, C.~J. P\'erez~Vicente, F.~Ritort, and
  R.~Spigler, ``The kuramoto model: A simple paradigm for synchronization
  phenomena,'' {\em Rev. Mod. Phys.}, vol.~77, pp.~137--185, Apr 2005.

\end{thebibliography}
\end{document}